\documentclass[conference]{IEEEtran}

\usepackage{graphicx}
\usepackage{amsmath,dsfont,bbm,epsfig,amssymb,amsfonts,  amstext, verbatim,amsopn,cite,subfigure,multirow,multicol,lipsum}
\usepackage{balance}
\usepackage{url}
\usepackage{amsfonts}
\usepackage{epsfig}
\usepackage{slashbox}
\usepackage{setspace}
\usepackage{stmaryrd}
\usepackage{psfrag}	
\usepackage{multirow}
\usepackage{float}
\usepackage{slashbox}
\usepackage[process=auto]{pstool}
\usepackage{etoolbox}
\usepackage{algorithmic}
\usepackage{algorithm}
\allowdisplaybreaks

\newcommand{\Equal}{\hspace{-0.4mm}=\hspace{-0.4mm}}
\newcommand{\Add}{\hspace{-0.4mm}+\hspace{-0.4mm}}
\newcommand{\Minus}{\hspace{-0.4mm}-\hspace{-0.4mm}}

\newtheorem{theo}{Theorem}
\newtheorem{lem}{Lemma}
\newtheorem{remk}{Remark}

\newtheorem{corol}{Corollary}

\newtoggle{Extended}

\toggletrue{Extended}

\iftoggle{Extended}{%

}{%
  
}

\EndPreamble
\begin{document}

\title{Resource Allocation for Mixed RF and Hybrid \\ RF/FSO Relaying \vspace{-0.3cm}}

\author{Vahid Jamali$^\dag$, Diomidis S. Michalopoulos$^\dag$, Murat Uysal$^\ddag$, and Robert Schober$^\dag$ \\
\IEEEauthorblockA{$^\dag$ Friedrich-Alexander University (FAU), Erlangen, Germany \hspace{0.4cm}
 $^\ddag$ Ozyegin University, Istanbul, Turkey \vspace{-0.5cm}}
}

\maketitle

\begin{abstract}
In this paper, we consider a mixed RF and hybrid RF/FSO system where several mobile users transmit their data over an RF link to a relay node (e.g. a small cell base station) and the relay forwards the information to a destination (e.g. a macro cell base station) over a hybrid RF/FSO backhaul link. The relay and the destination employ multiple antennas for  transmission and reception over the RF links while each mobile user has a single antenna. The RF links are full-duplex with respect to the FSO link and half-duplex with respect to each other, i.e., either the user-relay RF link or the relay-destination RF link is active. For this communication setup, we derive the optimal resource allocation policy for sharing the RF bandwidth resource between the RF links. Our numerical results show the effectiveness of the proposed communication architecture and resource allocation policy, and their superiority compared to existing schemes which employ only one type of backhaul link.
\end{abstract}


\section{Introduction} \label{Sec I (Intro)}

Free space optical (FSO) has recently re-emerged as an attractive option for wireless backhauling due to 
its large usable bandwidth  compared to traditional radio frequency (RF) backhauling \cite{UysalFSOsurvey}. Moreover, FSO transceivers employ very narrow beams which leads to several interesting features such as secure and interference-free  communication. However, these beneficial properties of FSO systems come at the expense of some drawbacks and challenges. One of the main challenges is the unpredictable connectivity of the FSO link due to atmospheric impairments such as atmospheric
turbulence and visibility limiting conditions including snow, fog, and dust \cite{UysalFSOsurvey}.

A possible strategy to mitigate the unpredictable connectivity of the  FSO link are so-called \textit{hybrid RF/FSO} links where an RF link is employed as a back-up for the FSO link \cite{HybridRFFSOShi,HybridRFFSORobert}. This option is attractive particularly due to the fact that distortions in the RF and FSO links are caused by different phenomena. Hence, a hybrid RF/FSO system is much more likely to maintain connectivity than a pure FSO system. FSO transceivers are mainly attractive for nodes with fixed locations where a line-of-sight link can be established. Motivated by this limitation of FSO for networks with mobile nodes, \textit{mixed RF/FSO} systems were proposed in the literature. Here, the RF and FSO links are cascaded, i.e., the mobile nodes employ RF links to send their information to an intermediate fixed node, i.e., a relay node, and the intermediate node forwards the information to the final destination via an FSO backhaul link  \cite{MixedRFFSOLett,MixedRFFSOAoluini,GeorgeFSO}. This communication setup can model several practical applications including: \textit{i)} Cellular communication where the mobile nodes   send their data to a relay station   and the relay station forwards the data to the base station; \textit{ii)} Femto cell networks where the mobile nodes in a building floor send their data to an access point and the access point forwards the information to the macro base station.  The performance  of mixed RF/FSO systems was investigated in  \cite{MixedRFFSOLett,MixedRFFSOAoluini} for a single-user amplify-and-forward (AF) relay network, and in  \cite{GeorgeFSO} for a multi-user decode-and-forward (DF) relay network.

In general,  the end-to-end performance of dual-hop communication is limited by the weakest link. Hence, in mixed RF/FSO systems, atmospheric turbulence may lead to a significant degradation of the end-to-end performance. Motivated by this limitation, in this paper, we consider a mixed RF and hybrid RF/FSO system where an additional RF backhaul link is employed as a back-up for the FSO backhaul link. Thereby, we assume that the back-up RF link for the relay-destination hop utilizes the same bandwidth resource as the RF link for the source-relay link. We consider a multi-user multiple-input multiple-output (MIMO) setup to fully exploit the available RF bandwidth. That is, relay and destination  are equipped with multiple antennas for data transmission over the RF links. For simplicity and feasibility reasons,  the relay is assumed to operate in the half-duplex mode with respect to  the RF links, i.e., it can either receive from the users or transmit to the destination over the RF links. Therefore,  we develop an optimal resource allocation policy which allocates the RF bandwidth resource to the two RF links based on the channel state information (CSI) of the RF and FSO links. The proposed protocol  adaptively switches between transmission and reception for the RF links which requires the relay to be equipped with a buffer \cite{Poor,NikolaJSAC} to temporarily store the data received from the users and forward it later over the FSO and/or RF backhaul links to the destination.   Our numerical results confirm the effectiveness of the proposed system architecture and resource allocation policy.

\textit{Notations:} We use the following notations throughout this paper:  $\mathbbmss{E}\{\cdot\}$ denotes expectation, $|\cdot|$ represents the magnitude of a complex number and the determinant of a matrix, $\angle$ denotes the phase of a complex number, and $\mathrm{erf}(\cdot)$ is the Gauss-error function.  Bold capital and small letters are used to denote matrices and vectors, respectively. $\mathbf{A}^H$ denotes the Hermitian transpose of $\mathbf{A}$, $\mathbf{I}_n$ is an $n\times n$ identity matrix, $\mathrm{diag}\{a_1,a_2,\dots,a_n\}$ is an $n\times n$ diagonal matrix whose diagonal elements are $a_1,a_2,\dots,a_n$, and $[\mathbf{A}]_{mn}$ denotes the element in the $m$-th row and $n$-th column of matrix $\mathbf{A}$. $\mathbf{1}\{\cdot\}\in\{0,1\}$ is an indicator function which is equal to one if the argument is true and equal to zero if it is not true. Additionally, $\mathrm{Rice}(\Omega,\Psi)$, $\mathrm{Unif}(a,b)$, and $\mathrm{GGamma}(\alpha,\beta)$ denote a Ricean random variable (RV) with parameters $\Omega$ and $\Psi$, a RV uniformly distributed in the interval $[a,b]$, and a Gamma-Gamma RV with parameters $\alpha$ and $\beta$, respectively.  For notational convenience, we use the definitions $ C(x) \triangleq \log_2(1+x)$, $[x]_{\mathrm{dB}}=10\log_{10}x$, and $\left[x\right]^+ \triangleq \max \{0,x\}$.

\section{Preliminaries and Assumptions}

In this section, we describe the considered system model, introduce the model for the mixed RF and hybrid RF/FSO communication links, and specify our assumptions regarding the CSI knowledge. 

\subsection{System Model}

\begin{figure}
\centering
\resizebox{0.9\linewidth}{!}{
\pstool[width=1.3\linewidth]{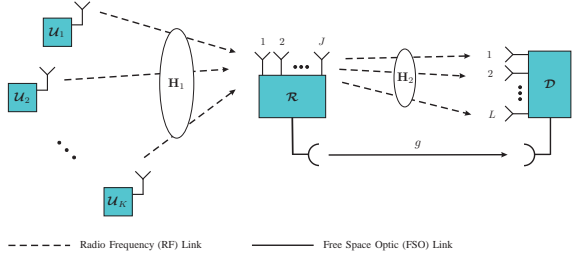}{
\psfrag{U1}[c][c][0.7]{$\boldsymbol{\mathcal{U}}_1$}
\psfrag{U2}[c][c][0.7]{$\boldsymbol{\mathcal{U}}_2$}
\psfrag{UK}[c][c][0.7]{$\boldsymbol{\mathcal{U}}_K$}
\psfrag{R}[c][c][0.7]{$\boldsymbol{\mathcal{R}}$}
\psfrag{D}[c][c][0.7]{$\boldsymbol{\mathcal{D}}$}
\psfrag{H1}[c][c][0.7]{$\mathbf{H}_1$}
\psfrag{H2}[c][c][0.7]{$\mathbf{H}_2$}
\psfrag{G}[c][c][0.7]{$g$}
\psfrag{N1}[c][c][0.6]{$1$}
\psfrag{N2}[c][c][0.6]{$2$}
\psfrag{Nn}[c][c][0.6]{$J$}
\psfrag{M1}[c][c][0.6]{$1$}
\psfrag{M2}[c][c][0.6]{$2$}
\psfrag{Mm}[c][c][0.6]{$L$}
\psfrag{RF}[l][c][0.6]{$\text{Radio Frequency (RF) Link}$}
\psfrag{FS}[l][c][0.6]{$\text{Free Space Optic (FSO) Link}$}
  } }\vspace{-0.4cm}
\caption{Proposed mixed system which consists of cascaded RF and hybrid RF/FSO links in a dual-hop configuration. The setup includes $K$ single-antenna users  $\boldsymbol{\mathcal{U}}_k,\,\,k=1,\dots,K$,  relay node  $\boldsymbol{\mathcal{R}}$, and    destination  $\boldsymbol{\mathcal{D}}$. \vspace{-0.4cm}}
\label{Fig:SysMod}
\end{figure}

The considered communication setup is schematically shown  in Fig. \ref{Fig:SysMod}. In particular, $K$ users $\boldsymbol{\mathcal{U}}_k,\,\,k=1,\dots,K$, wish to communicate with destination $\boldsymbol{\mathcal{D}}$ via relay node $\boldsymbol{\mathcal{R}}$.  There  is  no  direct communication link  between the users  and  the  destination, i.e.,  the  users can send their data to the destination only through the relay node.  There are two types of communication links in our system model: \textit{i)} the $\boldsymbol{\mathcal{U}}$-$\boldsymbol{\mathcal{R}}$ and $\boldsymbol{\mathcal{R}}$-$\boldsymbol{\mathcal{D}}$ RF links and \textit{ii)} the $\boldsymbol{\mathcal{R}}$-$\boldsymbol{\mathcal{D}}$ FSO link.  All RF links are assumed to use the same frequency. We assume that each user has a single antenna while the relay and the destination are equipped with $J\geq K$ and $L$ antennas, respectively. Furthermore, we consider the practical half-duplex constraint for the relay node, i.e., the relay cannot receive from the users through the $\boldsymbol{\mathcal{U}}$-$\boldsymbol{\mathcal{R}}$ RF fronthaul/access link and transmit to the destination via the $\boldsymbol{\mathcal{R}}$-$\boldsymbol{\mathcal{D}}$ RF backhaul link simultaneously. Since the FSO link does not interfere the RF links, the relay can always transmit data to the destination via the FSO backhaul link.

Let $T_{\boldsymbol{\mathcal{UR}}}$, $T_{\boldsymbol{\mathcal{RD}}}^{\mathrm{RF}}$, and $T_{\boldsymbol{\mathcal{RD}}}^{\mathrm{FSO}}$ denote the coherence time of the $\boldsymbol{\mathcal{U}}$-$\boldsymbol{\mathcal{R}}$ RF link, the $\boldsymbol{\mathcal{R}}$-$\boldsymbol{\mathcal{D}}$ RF link, and the $\boldsymbol{\mathcal{R}}$-$\boldsymbol{\mathcal{D}}$ FSO link, respectively. Since the users are mobile nodes whereas both the relay and the destination are fixed nodes, the $\boldsymbol{\mathcal{U}}$-$\boldsymbol{\mathcal{R}}$ RF link fluctuates considerably faster than the  $\boldsymbol{\mathcal{R}}$-$\boldsymbol{\mathcal{D}}$ RF link and the FSO link. Hence, we assume that $T_{\boldsymbol{\mathcal{UR}}} \ll T_{\boldsymbol{\mathcal{RD}}}^{\mathrm{RF}},T_{\boldsymbol{\mathcal{RD}}}^{\mathrm{FSO}}$ holds. Thereby, we consider the coherence time of the $\boldsymbol{\mathcal{U}}$-$\boldsymbol{\mathcal{R}}$ RF link as the time reference for resource allocation. In particular, we assume the entire time of operation is divided into $B$ blocks of length $T_{\boldsymbol{\mathcal{UR}}}$ where each block consists of $N$ symbol intervals for the RF signal and each node transmits a codeword which spans one block or a fraction of one block. Furthermore, we assume that the FSO link has a larger bandwidth than the RF links. In order to model this, we assume that the duration of the symbol intervals of the FSO signal is $M$ times smaller than that of the RF signals, i.e., the FSO signal has a symbol rate that is $M$ times larger than the symbol rate of the RF signal.  Additionally, we  assume  that  users  always  have  enough information  to  send  in  all blocks  and  that  the  number
of  blocks satisfies $B\to\infty$. 

\subsection{Communication Links}

In the following, we describe the channel models for the RF and FSO links  which are used throughout the paper. 


\noindent
\textbf{RF Links:} The $\boldsymbol{\mathcal{U}}$-$\boldsymbol{\mathcal{R}}$ RF link constitutes a $(K,J)$-distributed MIMO system with $K$ single-antenna transmitters and one $J$-antenna receiver. In contrast, the $\boldsymbol{\mathcal{R}}$-$\boldsymbol{\mathcal{D}}$ RF backhaul link is a standard $(J,L)$-point-to-point MIMO system. The received  signals at the relay and the destination can be modelled as 
\begin{IEEEeqnarray}{lll}\label{Eq:GaussianRF}
  \mathbf{y}_j^n [b]  =  \mathbf{H}_j[b] \mathbf{x}_j^n [b] + \mathbf{z}_j^n [b], \quad j=1,2
\end{IEEEeqnarray}
where $\mathbf{x}_1^n[b]\in\mathbb{C}^{K\times 1}$ and $\mathbf{x}_2^n[b]\in\mathbb{C}^{J\times 1}$ denote the transmit symbol vectors of the users and the relay, respectively,  $\mathbf{y}_1^n[b]\in\mathbb{C}^{J\times 1}$ and $\mathbf{y}_2^n[b]\in\mathbb{C}^{L\times 1}$ are the received  symbol vectors  at the relay and the destination, respectively, and  $\mathbf{z}_1^n[b]\in\mathbb{C}^{J\times 1}$ and $\mathbf{z}_2^n[b]\in\mathbb{C}^{L\times 1}$  denote the noise vectors  at  the relay and  the destination, respectively, in the $n$-th symbol interval of the $b$-th block.  We  assume  that   $\mathbf{z}_1^n [b]$ and $\mathbf{z}_2^n [b]$ are zero-mean complex additive white Gaussian noise (AWGN) vectors with covariance matrices $\sigma_{\boldsymbol{\mathcal{R}}}^2\mathbf{I}_{J}$ and $\sigma_{\boldsymbol{\mathcal{D}}}^2\mathbf{I}_{L}$, respectively, and are mutually  independent  and independent from  the  transmitted codewords. The variance of the noise at the RF receivers   is  given  by  $[\sigma_{j}^2]_{\mathrm{dB}}=W^{\mathrm{RF}}N_{j,0}+N_{j,F},\,\,j\in\{\boldsymbol{\mathcal{R}},\boldsymbol{\mathcal{D}}\}$,  where $W^{\mathrm{RF}}$, $N_{j,0}$, and $N_{j,F}$ denote the RF bandwidth, the noise power
spectral density (in dB/Hz), and the noise figure of the
receiver, respectively. Furthermore, $\mathbf{H}_1[b]\in\mathbb{C}^{J \times K}$ and $\mathbf{H}_2[b]\in\mathbb{C}^{L \times J}$ denote the channel coefficient matrices of the $\boldsymbol{\mathcal{U}}$-$\boldsymbol{\mathcal{R}}$ and $\boldsymbol{\mathcal{R}}$-$\boldsymbol{\mathcal{D}}$ RF links, respectively.  Moreover,  we assume that all  elements of   $\mathbf{H}_1[b]$ and $\mathbf{H}_2[b]$ are mutually  independent,  ergodic,  and stationary  random  processes with  continuous  probability  density  functions. Throughout the paper, we assume the RF links are affected by Ricean fading \cite{WiMAXFSO,HybridRFFSORobert}. We further take into account the effect of path loss. Suppose $h[b]$ is one of the elements of $\mathbf{H}_j[b],\,\,j=1,2$. Thereby, $h[b]$ can be written as $h[b]= \sqrt{h_{a}} h_{f}[b]$, where  $h_{a}$ and $h_{f}[b]$ are the average power gain  and the fading coefficient of the RF link which are given by \cite{WiMAXFSO,HybridRFFSORobert}
\begin{IEEEeqnarray}{lll}\label{Eq:ModelRF}
 \begin{cases}
 h_a = \left[\frac{\lambda^{\mathrm{RF}}\sqrt{G^{\mathrm{RF}}_t G^{\mathrm{RF}}_r}}{4\pi d^{\mathrm{RF}}_{\mathrm{ref}}}\right]^2 \times \left[\frac{d^{\mathrm{RF}}_{\mathrm{ref}}}{d^{\mathrm{RF}}}\right]^\nu \\
  |h_f[b]|  \sim \mathrm{Rice}(\Omega,\Psi),\,\, \angle h_f[b] \sim \mathrm{Unif}(-\pi,\pi)
 \end{cases}
\end{IEEEeqnarray}
where $\lambda^{\mathrm{RF}}$ is the wavelength of the RF signal, $G^{\mathrm{RF}}_t$ and $G^{\mathrm{RF}}_r$ are the RF transmit and receive
antenna gains, respectively, $d^{\mathrm{RF}}_{\mathrm{ref}}$ is a reference distance for the antenna far-field, $d^{\mathrm{RF}}$ is the distance between RF transmitter and receiver, and $\nu$ is the RF path-loss exponent. Parameters $\Omega$ and $\Psi$ in the Rice  distribution are the ratio between the power in the direct path and the power in the scattered paths and  the total power in both paths, respectively. For the $\boldsymbol{\mathcal{U}}$-$\boldsymbol{\mathcal{R}}$ RF link, we assume that the direct link is not available ($\Omega=0$). Hence, the Ricean fading reduces to Rayleigh fading. 
Finally, the transmitted RF signals have to meet the following power constraints
\begin{IEEEeqnarray}{lll}\label{Eq:PowerRF}
  \mathbbmss{E}\{ \mathbf{x}_1^n [b]  (\mathbf{x}_1^n [b])^H \} \leq \mathrm{diag} \big\{P_{\boldsymbol{\mathcal{U}}}^{1},P_{\boldsymbol{\mathcal{U}}}^{2},\dots,P_{\boldsymbol{\mathcal{U}}}^{K} \big\} \IEEEyesnumber\IEEEyessubnumber \\
  \mathbbmss{E}\{ (\mathbf{x}_2^n [b])^H  \mathbf{x}_2^n [b] \} \leq P_{\boldsymbol{\mathcal{R}}}^{\mathrm{RF}},  \IEEEyessubnumber 
\end{IEEEeqnarray}
where $P_{\boldsymbol{\mathcal{U}}}^{k}$ and $P_{\boldsymbol{\mathcal{R}}}^{\mathrm{RF}}$ are the RF transmit powers of user $k$ and the relay, respectively.

\noindent
\textbf{FSO Link:}  The relay  node  is
equipped with an  aperture transmitter pointing in the direction of a photo detector at the destination. We assume an  intensity  modulation  direct detection  (IM/DD)  FSO  system with  on-off keying (OOK) modulation for the FSO  link, i.e., the photo detector directly detects the intensity of the received photo-current by integrating over each symbol interval. In particular, after removing the ambient background light intensity, the detected signal intensity at the destination is modelled as \cite{HybridRFFSORobert}
\begin{IEEEeqnarray}{lll}\label{Eq:GaussianFSO}
  y^m[b]  = \rho g[b] x^m[b] + z^m[b], 
\end{IEEEeqnarray}
where $\rho$ is the responsivity of the photo detector and $x^m[b]\in \big\{0,P_{\boldsymbol{\mathcal{R}}}^{\mathrm{FSO}}\big\}$, $y^m[b]\in\mathbb{R}$, and $z^m[b]\in\mathbb{R}$ are the power of the OOK modulated symbol at the relay, the detected signal power at the destination, and the   shot noise  at the destination caused by ambient light for the $m$-th symbol of the $b$-th block, respectively. Moreover, $P_{\boldsymbol{\mathcal{R}}}^{\mathrm{FSO}}$ is the maximum allowable transmit power of the relay over the FSO link which is mainly determined by  eye safety regulations \cite{UysalFSOsurvey}. Noise  $z^m [b]$ is modelled as zero-mean real AWGN with variance  $\sigma^2$  and is independent  of  the  transmitted signal. Moreover, $g[b]$ is modelled as $g[b]=  g_{a} g_{f}[b]$, where $g_{a}$ and $g_{f}[b]$ are the average gain and the fading gain of the FSO link, respectively, and are given by \cite{HybridRFFSOShi}
\begin{IEEEeqnarray}{lll}\label{Eq:ModelFSO}
 \begin{cases}
 g_a  = \left[\mathrm{erf}\left(\frac{\sqrt{\pi} r}{\sqrt{2}  \phi d^{\mathrm{FSO}}}\right)\right]^2 \times 10^{-\kappa d^{\mathrm{FSO}}/10} \\
g_f[b] \sim \mathrm{GGamma}(\alpha,\beta)
 \end{cases}
\end{IEEEeqnarray}
where $r$ is the aperture radius,  $\phi$ is the divergence angle of the beam, $d^{\mathrm{FSO}}$ is the distance between  FSO transmitter and receiver, and $\kappa$ is the weather-dependent attenuation coefficient. Assuming spherical wave propagation, parameters $\alpha$ and $\beta$ in the Gamma-Gamma distribution are related to physical parameters as follows \cite{HybridRFFSORobert}
\begin{IEEEeqnarray}{lll}\label{Eq:AlpBet}
\alpha = \left[   \exp\left\{\frac{0.49 \vartheta^2 }{\left[1+0.18 \xi^2 + 0.56 \vartheta^{12/5}\right]^{7/6}}\right\}-1\right]^{-1} \IEEEyesnumber\IEEEyessubnumber \\
 \beta = \left[   \exp\left\{\frac{0.51 \vartheta^2 \left[ 1+0.69 \vartheta^{12/5}\right]^{-5/6} }{\left(1+0.9 \xi^2 + 0.62 \xi^2 \vartheta^{12/5}\right)^{5/6}}\right\}-1\right]^{-1}, \IEEEyessubnumber \quad
\end{IEEEeqnarray}
where $\vartheta^2=0.5C_n^2\varsigma^{7/6} (d^{\mathrm{FSO}})^{11/6}$, $\xi^2=\varsigma r^2/d^{\mathrm{FSO}}$, and $\varsigma=2\pi/\lambda^{\mathrm{FSO}}$. Here, $\lambda^{\mathrm{FSO}}$ is the wavelength and $C_n^2$  is the weather-dependent index of refraction structure parameter.

\subsection{CSI Knowledge}

Throughout this paper, we assume that the relay has full knowledge of the CSI of all links and is responsible for determining the optimal transmission strategy
and for conveying the strategy to the other nodes, cf. Theorem \ref{Theo:Opt_Prot}. Moreover, since for large $K$, an excessive amount of CSI feedback is required for adaptive rate transmission at the user nodes, we assume that the users transmit with a priori  fixed transmission rates and hence, no CSI knowledge is required at the users. We note, however, that the proposed problem formulation and the resulting protocol are provided in a general form and the case where the users transmit with adaptive rates can be easily accommodated, cf. Remark~\ref{Remk:Cap_General}. Furthermore, the destination knows the CSI  of the $\boldsymbol{\mathcal{R}}$-$\boldsymbol{\mathcal{D}}$ links as required for reliable coherent decoding. Moreover, we assume that the channel states change slowly enough such that the signaling overhead caused by channel estimation and feedback is negligible compared to the amount of  information transmitted in one block.

\section{Mixed RF and Hybrid RF/FSO Protocol}

In this section, we first introduce the considered protocol and explain the adopted coding scheme. Subsequently, we derive the optimal resource allocation policy for the proposed protocol.

\subsection{The Proposed Protocol}

Recall that due to the half-duplex constraint, the relay cannot use the $\boldsymbol{\mathcal{U}}$-$\boldsymbol{\mathcal{R}}$ and $\boldsymbol{\mathcal{R}}$-$\boldsymbol{\mathcal{D}}$ RF links simultaneously. In fact, if the quality of the FSO backhaul link is sufficiently good, all data received  at the relay from the users can be forwarded using the FSO link to the destination, and there is no need to activate the $\boldsymbol{\mathcal{R}}$-$\boldsymbol{\mathcal{D}}$ RF backhaul link at all. However, due to the time-varying fluctuations of the FSO link, the relay may not always be able to forward the users' information to the destination by employing only the FSO link. Hence,  the $\boldsymbol{\mathcal{R}}$-$\boldsymbol{\mathcal{D}}$ RF link is needed as a back-up. On the other hand, using the   $\boldsymbol{\mathcal{R}}$-$\boldsymbol{\mathcal{D}}$ RF backhaul link comes at the expense of reducing the time of transmission of the users to the relay  over the $\boldsymbol{\mathcal{U}}$-$\boldsymbol{\mathcal{R}}$ RF link  due to the half-duplex constraint. 

In light of the above discussion, the main idea of the proposed protocol is to adaptively share the RF bandwidth resource between the $\boldsymbol{\mathcal{U}}$-$\boldsymbol{\mathcal{R}}$  RF fronthaul and  the $\boldsymbol{\mathcal{R}}$-$\boldsymbol{\mathcal{D}}$ RF backhaul link  based on the fading state in the $b$-th block, i.e., $\mathbf{H}_1[b]$, $\mathbf{H}_2[b]$, and $g[b]$. To this end, we introduce time sharing variable $q[b]\in[0, 1]$ where $q[b]$ denotes the fraction of block $b$ in which the relay receives, i.e., the $\boldsymbol{\mathcal{U}}$-$\boldsymbol{\mathcal{R}}$ RF link is active. In the remaining $1-q[b]$ fraction of block $b$, the relay transmits over the $\boldsymbol{\mathcal{R}}$-$\boldsymbol{\mathcal{D}}$ RF link. We assume that the relay node is equipped with an infinite-size buffer for data storage. Let $Q[b]$ denote the number of information bits available in the buffer after receiving  the information from the users in the  $b$-th  block. Using these notations, the coding schemes, the transmission
rates, and the dynamics of the queues at the buffers are presented below.

\noindent
\textbf{$\boldsymbol{\mathcal{U}}$-$\boldsymbol{\mathcal{R}}$ RF Link:}  The users employ Gaussian codebooks, i.e., the $k$-th element of vector $\mathbf{x}_1^n[b]$ is generated independently according to a zero-mean rotationally invariant complex Gaussian distribution with variance $P_{\boldsymbol{\mathcal{U}}}^{k}$. At the beginning of each fading block,  each user $k$ encodes $q[b]NR_{\boldsymbol{\mathcal{U}}}^{k}$ bits of information into a codeword taken from a Gaussian codebook with a fixed rate $R_{\boldsymbol{\mathcal{U}}}^{k}$ bits/symbol and having a length of $q[b]N$ symbols. The users transmit their codewords and the relay receives $\mathbf{y}_1^n [b]$ according to (\ref{Eq:GaussianRF}). The relay can employ several multi-user detection schemes proposed in the literature \cite{GeorgeFSO}, e.g.,  linear zero-forcing (ZF) and minimum mean square error (MMSE)
 detection, nonlinear detection schemes incorporating  successive interference cancellation (SIC), or optimal maximum likelihood (ML) detection. Depending on the type of detector used, the codeword of each user experiences a certain signal-to-noise ratio (SNR) denoted by $\gamma_{\boldsymbol{\mathcal{U}}}^{k}[b]$. For instance, for ZF detection, $\gamma_{\boldsymbol{\mathcal{U}}}^{k}[b]$ is given by
\begin{IEEEeqnarray}{lll}\label{Eq:ZF_SNR}
  \gamma_{\boldsymbol{\mathcal{U}}}^{k}[b] = \frac{P_{\boldsymbol{\mathcal{U}}}^{k}}{\sigma_{\boldsymbol{\mathcal{R}}}^2 \left[\big((\mathbf{H}_1[b])^H\mathbf{H}_1[b]\big)^{-1} \right]_{kk}}.
\end{IEEEeqnarray}
The codeword of user $k$ can be decoded reliably only if  $\gamma_{\boldsymbol{\mathcal{U}}}^{k}[b]\geq 2^{R_{\boldsymbol{\mathcal{U}}}^{k}}-1$; otherwise  the relay cannot decode this codeword and has to drop it and ask user $k$ to retransmit this information in the following blocks. The effective rate of the $\boldsymbol{\mathcal{U}}$-$\boldsymbol{\mathcal{R}}$ RF link is defined as the normalized sum of the information bits in bits/symbol  that can be decoded successfully  at the relay and is given by 
\begin{IEEEeqnarray}{lll}\label{Eq:UR_Rate_RF}
  C_1^{\mathrm{RF}}[b] = \sum_{k=1}^{K}  \mathbf{1}\big\{\gamma_{\boldsymbol{\mathcal{U}}}^{k}[b]\geq 2^{R_{\boldsymbol{\mathcal{U}}}^{k}}-1\big\} R_{\boldsymbol{\mathcal{U}}}^{k}.
\end{IEEEeqnarray}

\noindent
\textbf{$\boldsymbol{\mathcal{R}}$-$\boldsymbol{\mathcal{D}}$ RF Link:} This backhaul link is a standard point-to-point MIMO channel with an average power constraint across all antennas. We employ Gaussian codebooks, i.e.,   vector $\mathbf{x}_2^n[b]$ is a multivariate zero-mean rotationally invariant complex Gaussian vector, and  waterfilling power allocation across the transmit antennas \cite{MIMOTelatar}. We note that the relay cannot transmit more information than what it has  stored in its buffer and has not yet been  transmitted over the FSO link. Thereby, the relay extracts $\min\big\{Q[b],[1-q[b]] N C_2^{\mathrm{RF}}[b]\big\}$ bits of information from its buffer and transmits them over its $J$ antennas where
\begin{IEEEeqnarray}{lll}\label{Eq:RU_Cap_RF}
  C_2^{\mathrm{RF}}[b] =  \sum_{j=1}^{\min\{J,L\}} \bigg[\log_2\bigg\{ \mu \frac{\chi_j^2}{\sigma_{\boldsymbol{\mathcal{D}}}^2}\bigg\}\bigg]^+.
\end{IEEEeqnarray}
Here, $\chi_j$ is the $j$-th singular value of $\mathbf{H}_2$ \cite{MIMOTelatar}; $\mu$ is the water level which is chosen to satisfy the power constraint in (\ref{Eq:PowerRF}b) as the solution of the following equation
\begin{IEEEeqnarray}{lll}
\sum_{j=1}^{\min\{J,L\}} \bigg[ \mu - \frac{\sigma_{\boldsymbol{\mathcal{D}}}^2}{\chi_j^2}\bigg]^+ = P_{\boldsymbol{\mathcal{R}}}^{\mathrm{RF}}.
\end{IEEEeqnarray}

\noindent
\textbf{FSO Link:} We employ OOK  modulation and soft detection for the FSO  link. From an information theoretical point of view, the FSO link can be modelled as a binary input-continuous output  AWGN channel where the capacity is achieved with a uniform distribution of the binary inputs \cite{UysalFSOCap,OOKCapacity}. Thereby, at the beginning of each block, the relay extracts $\min\big\{[Q[b-1]-(1-q[b-1])N C_2^{\mathrm{RF}}[b-1]]^+,MN C^{\mathrm{FSO}}[b]\big\}$ bits of information from its buffer, encodes them into a codeword, and sends the codeword to the destination where $C^{\mathrm{FSO}}[b]$ is the capacity of the Gaussian channel with OOK input and is given by \cite{UysalFSOCap,OOKCapacity}
\begin{IEEEeqnarray}{lll}\label{Eq:RU_Cap_FSO}
  C^{\mathrm{FSO}}[b] = 1 \Minus \frac{1}{2\sqrt{\pi}} \int_{-\infty}^{+\infty}  \hspace{-0.3cm} \exp\left\{-t^2\right\} 
  \log_2\bigg\{1 \Add \exp\left\{-\frac{p^2[b]}{2\sigma^2}\right\} \nonumber \\
  \left[ \exp\left\{\frac{2tp[b]}{\sqrt{2\sigma^2}}\right\} \Add  \exp\left\{-\frac{2tp[b]}{\sqrt{2\sigma^2}}\right\} \Add \exp\left\{-\frac{p^2[b]}{2\sigma^2}\right\}  \right]\bigg\} \mathrm{d}t,\quad\quad
\end{IEEEeqnarray}
where $p[b]=\rho g[b] P_{\boldsymbol{\mathcal{R}}}^{\mathrm{FSO}}$.
%

\noindent
\textbf{Dynamics of the Queue:}  After the relay has received the information from the users in the $b$-th block,  the amount of information bits in the buffer is updated as
\begin{IEEEeqnarray}{lll}\label{Eq:UR_Q_RF}
  Q[b]=\Big[Q[b-1]&-(1-q[b-1])NC_2^{\mathrm{RF}}[b-1] \nonumber \\
  &- MN C^{\mathrm{FSO}}[b] \Big]^+ + q[b]N C_1^{\mathrm{RF}}[b]. \quad
\end{IEEEeqnarray}

\subsection{Optimal Resource Allocation}

In this section our goal is to obtain the optimal resource allocation for the RF links, i.e., the optimal $q[b],\,\,\forall b$, such that the average number of information bits received at the destination, denoted by $\tau$, is maximized. Note that since there is no direct link, the amount of information received from the users at the destination is identical to that received from the relay at the destination. Hence, the throughput maximization problem can be formulated as 
\begin{IEEEeqnarray}{lll} \label{Eq:Prob_Delay_Tol} 
\underset{q[b]\in[0,1],\,\,\forall b}{\mathrm{maximize}}  \tau \Equal \hspace{-0.5mm} \underset{B\to\infty}{\lim}\frac{1}{B} \hspace{-0.5mm} \sum_{b=1}^{B} \hspace{-0.5mm} \min\big\{Q[b],(1\Minus q[b]) N C_2^{\mathrm{RF}}[b]\big\} \,\,\,\, \\
 \Add\min\big\{\hspace{-0.05cm}[Q[b\Minus 1]\Minus(1\Minus q[b\Minus 1])N C_2^{\mathrm{RF}}[b\Minus 1]]^+\hspace{-0.1cm}, MN C^{\mathrm{FSO}}[b]\hspace{-0.05cm}\big\}.\nonumber
\end{IEEEeqnarray}
Finding the maximum throughput and the corresponding optimal resource allocation policy is challenging for the considered channel model due to the recursive dynamics of the queue. Nevertheless, we can obtain an upper bound on the achievable throughput of the proposed protocol by neglecting the effect of the queues on the transmission rate in (\ref{Eq:Prob_Delay_Tol}). 

\textit{Upper Bound $\tau^{\mathrm{upp}}$:} The achievable average rate of the proposed protocol for the considered mixed RF and hybrid RF/FSO system with block fading is upper bounded by $\tau^{\mathrm{upp}}$ obtained from the following optimization problem
\begin{IEEEeqnarray}{llll} \label{Eq:ProbUpp}
\underset{q[b]\in[0,1],\,\,\forall b}{\mathrm{maximize}}\quad  \tau^{\mathrm{upp}}  \quad 
\mathrm{subject \,\, to} \vspace{-0.2cm} \\
\mathrm{C1:}\,\tau^{\mathrm{upp}} \leq  \underset{B\to\infty}{\lim} \frac{1}{B} \sum_{b=1}^{B} q[b] NC_1^{\mathrm{RF}}[b] \nonumber \\
\mathrm{C2:}\, \tau^{\mathrm{upp}} \leq  \underset{B\to\infty}{\lim} \frac{1}{B} \sum_{b=1}^{B}\big[(1-q[b])NC_2^{\mathrm{RF}}[b] + NMC^{\mathrm{FSO}}[b] \big ]. \nonumber
\end{IEEEeqnarray}

In the following theorem, we provide the optimal solution to the above optimization problem. For notational simplicity, let $C_1^{\mathrm{RF}}(\mathbf{H}_1)$, $C_2^{\mathrm{RF}}(\mathbf{H}_2)$, and $C^{\mathrm{FSO}}(g)$ denote the capacity functions given in (\ref{Eq:UR_Rate_RF}), (\ref{Eq:RU_Cap_RF}), and (\ref{Eq:RU_Cap_FSO}) in terms of the fading states $\mathbf{H}_1$, $\mathbf{H}_2$, and $g$, respectively. Moreover, let $f_{\mathbf{H}_1}(\mathbf{H}_1)$, $f_{\mathbf{H}_2}(\mathbf{H}_2)$, and $f_{g}(g)$ denote the probability density functions of $\mathbf{H}_1$, $\mathbf{H}_2$, and $g$, respectively.

\begin{theo}\label{Theo:Opt_Prot}
The optimal resource allocation policy depends only on the values of channel matrices $\mathbf{H}_1$ and $\mathbf{H}_2$ and is denoted by  $q^*(\mathbf{H}_1,\mathbf{H}_2)\in[0,1]$. Moreover, the optimal resource allocation policy reduces to the adaptive activation of either the $\boldsymbol{\mathcal{U}}$-$\boldsymbol{\mathcal{R}}$ link or the $\boldsymbol{\mathcal{R}}$-$\boldsymbol{\mathcal{D}}$ RF link according to
\begin{IEEEeqnarray}{lll}\label{Eq:Opt_Prot}
 q^*(\mathbf{H}_1,\mathbf{H}_2) =\begin{cases}
 0,\quad &\mathrm{if} \,\, \lambda C_1^{\mathrm{RF}}(\mathbf{H}_1) \geq [1-\lambda]C_2^{\mathrm{RF}}(\mathbf{H}_2)  \\
 1,\quad &\mathrm{otherwise}
 \end{cases}\quad
\end{IEEEeqnarray}
where $\lambda\in (0,1]$ is a constant which does not depend on the instantaneous realization of the fading but depends on the fading distributions. The optimal value of $\lambda$ can be obtained offline before transmission starts using an iterative algorithm   with  the following update equation 
\begin{IEEEeqnarray}{lll}\label{Eq:Update}
 \lambda[i+1] =  \Big[\lambda[i]-\delta[i]\big[\bar{C}_1^{\mathrm{RF}}[i]  - \bar{C}_2^{\mathrm{RF}}[i] -  M \bar{C}^{\mathrm{FSO}} \big] \Big]_0^1,\qquad
\end{IEEEeqnarray}
where $i$ is the iteration index and $\delta[i]$ is an appropriately chosen step size parameter. The average capacity rates $\bar{C}_1^{\mathrm{RF}}[i]$, $\bar{C}_2^{\mathrm{RF}}[i]$, and $\bar{C}^{\mathrm{FSO}}$ are given by
\begin{IEEEeqnarray}{rll}\label{Eq:AlgCap}
  \bar{C}_1^{\mathrm{RF}}[i] \,\,&= \mathbbmss{E}\big\{ q^*(\mathbf{H}_1,\mathbf{H}_2) C_1^{\mathrm{RF}}(\mathbf{H}_1) \big\} \nonumber \\
  &= \iint    q^*(\mathbf{H}_1,\mathbf{H}_2) C_1^{\mathrm{RF}}(\mathbf{H}_1) \nonumber \\
  &\qquad \times f_{\mathbf{H}_1}(\mathbf{H}_1) f_{\mathbf{H}_2}(\mathbf{H}_2)  \mathrm{d}\mathbf{H}_1 \mathrm{d}\mathbf{H}_2 \qquad \IEEEyesnumber \IEEEyessubnumber \\  
  \bar{C}_2^{\mathrm{RF}}[i] \,\,&= \mathbbmss{E}\big\{[1-q^*(\mathbf{H}_1,\mathbf{H}_2)] C_2^{\mathrm{RF}}(\mathbf{H}_2) \big\} \nonumber \\
  & = \iint  [1- q^*(\mathbf{H}_1,\mathbf{H}_2)] C_2^{\mathrm{RF}}[\mathbf{H}_2] \nonumber \\%
  &\qquad \times f_{\mathbf{H}_1}(\mathbf{H}_1) f_{\mathbf{H}_2}(\mathbf{H}_2)  \mathrm{d}\mathbf{H}_1 \mathrm{d}\mathbf{H}_2  \IEEEyessubnumber \\ 
  \bar{C}^{\mathrm{FSO}}  \,\,&= \mathbbmss{E}\big\{  C^{\mathrm{FSO}}\big\}= \int   C^{\mathrm{FSO}}(g)  f_{g}(g) \mathrm{d}g,  \IEEEyessubnumber
\end{IEEEeqnarray}
where for a given $\lambda=\lambda[i]$ in the $i$-th iteration, $q^*(\mathbf{H}_1,\mathbf{H}_2)$ has to be obtained from (\ref{Eq:Opt_Prot}) for a given set of fading values. 
Employing the optimal $\lambda^*$ obtained from the iterative algorithm  and the optimal   $q^*(\mathbf{H}_1,\mathbf{H}_2)$,  from (\ref{Eq:Opt_Prot}), and substituting them into (\ref{Eq:AlgCap}), the upper bound $\tau^{\mathrm{upp}}$ is obtained as
\begin{IEEEeqnarray}{lll}\label{Eq:OptUpp}
  \tau^{\mathrm{upp}} = N \min\big\{\bar{C}_1^{\mathrm{RF}},\bar{C}_2^{\mathrm{RF}}+M\bar{C}^{\mathrm{FSO}}\big\}.  
\end{IEEEeqnarray}

\end{theo}
\begin{IEEEproof}
\iftoggle{Extended}{%
   Please refer to Appendix \ref{App:Theo_Opt_Prot}. 
}{%
  Due to the space constraint, we provide the detailed proof
in \cite[Appendix A]{GlobeCOM2015Arxiv} which is an extended version of this paper. Here, we provide only a sketch of the proof. The  standard  Lagrange  method  is  employed  to  solve
the  optimization  problem  in  (\ref{Eq:ProbUpp})  \cite{Boyd}.  In  particular,  let $\lambda_1$ and $\lambda_2$ denote
the  Lagrange  multipliers  corresponding  to    constraints $\mathrm{C1}$ and $\mathrm{C2}$  in
(\ref{Eq:ProbUpp}).  After  forming  the  Lagrangian  function,  the  optimal
resource allocation  for  given $\lambda_1$ and $\lambda_2$
is obtained by equating the derivative of the Lagrangian
function  with  respect  to  $q[b]$ to zero. Moreover, we find that $\lambda_1+\lambda_2=1$ has to hold. Hence, we define $\lambda_1=\lambda$ and $\lambda_2=1-\lambda$. Then, the optimal $\lambda$ can be found with the well-known gradient  method with update equation  (\ref{Eq:Update}). Finally, the optimal $\tau^{\mathrm{upp}}$ is the minimum of the two bounds given in constraints $\mathrm{C1}$ and $\mathrm{C2}$ after substituting $q^*(\mathbf{H}_1,\mathbf{H}_2)$ from (\ref{Eq:Opt_Prot}) which leads to (\ref{Eq:OptUpp}).
}
\end{IEEEproof}

In the following, we formally state the achievability of the upper bound $\tau^{\mathrm{upp}}$ as $B\to\infty$.

\begin{lem}\label{Lem:Achiev}
The upper bound $\tau^{\mathrm{upp}}$ is achievable if the optimal resource allocation policy in Theorem~\ref{Theo:Opt_Prot} is employed. More precisely, as $B\to\infty$, we obtain $\tau\to\tau^{\mathrm{upp}}$.
\end{lem}
\begin{IEEEproof}
\iftoggle{Extended}{%
Please refer to Appendix \ref{App:Lem_Achiev}.
}{%
  The proof follows a well-known property in queuing theory \cite{Neely}. The main idea is that although the \textit{instantaneous} departure rate of a queue is limited by the amount of information in the queue, the \textit{average} departure rate (over time as $B\to\infty$)  is limited only by the average arrival rate and the average queue process which correspond to $N\bar{C}_1^{\mathrm{RF}}$ and $N\bar{C}_2^{\mathrm{RF}}+MN\bar{C}^{\mathrm{FSO}}$, respectively, for our system model. This bound on the average throughput, $\tau$, is identical to the one for  $\tau^{\mathrm{upp}}$ in (\ref{Eq:OptUpp}). A more elaborate proof is provided in \cite[Appendix B]{GlobeCOM2015Arxiv}.
}
\end{IEEEproof}

We note that Lemma~\ref{Lem:Achiev} reveals that $\tau^{\mathrm{upp}}$ is \textit{asymptotically} achievable by using the resource allocation policy in Theorem~\ref{Theo:Opt_Prot}. In fact, it is possible that for some fading blocks the relay does not have enough information in its queue to send to the destination over the  FSO and $\boldsymbol{\mathcal{R}}$-$\boldsymbol{\mathcal{D}}$ RF links, but as $B\to\infty$, the effect of these events on the throughput becomes negligible, i.e., $\tau\to\tau^{\mathrm{upp}}$.

\begin{corol}\label{Corol:Analytic}
The optimal value of Lagrange multiplier $\lambda$ can be obtained analytically from one of  the following mutually exclusive cases:

\noindent
\textit{Case 1:} If $\mathbbmss{E}\big\{C_1^{\mathrm{RF}}(\mathbf{H}_1) \big\} \leq M \mathbbmss{E}\big\{C^{\mathrm{FSO}}(g) \big\}$ holds, we obtain $\lambda^*=1$. Hence, we have $q^*(\mathbf{H}_1,\mathbf{H}_2)=1,\,\,\forall  \mathbf{H}_1,\mathbf{H}_2$.

\noindent
\textit{Case 2:} If $\mathbbmss{E}\big\{C_1^{\mathrm{RF}}(\mathbf{H}_1) \big\} >  M \mathbbmss{E}\big\{C^{\mathrm{FSO}}(g) \big\}$ holds, the optimal value of $\lambda$ is obtained analytically from the following equation 
\begin{IEEEeqnarray}{lll}\label{Eq:Eq_Opt_Lam}
\iint   q^*(\mathbf{H}_1,\mathbf{H}_2) C_1^{\mathrm{RF}}(\mathbf{H}_1)  f_{\mathbf{H}_1}(\mathbf{H}_1) f_{\mathbf{H}_2}(\mathbf{H}_2)  \mathrm{d}\mathbf{H}_1 \mathrm{d}\mathbf{H}_2 
 \nonumber \\
= \iint  [1-q^*(\mathbf{H}_1,\mathbf{H}_2)]   C_2^{\mathrm{RF}}(\mathbf{H}_2)  f_{\mathbf{H}_1}(\mathbf{H}_1) f_{\mathbf{H}_2}(\mathbf{H}_2)  \mathrm{d}\mathbf{H}_1 \mathrm{d}\mathbf{H}_2
\nonumber \\ \quad + M \int   C^{\mathrm{FSO}}(g)  f_{g}(g) \mathrm{d}g, 
\end{IEEEeqnarray}
where $q^*(\mathbf{H}_1,\mathbf{H}_2)$ is substituted from (\ref{Eq:Opt_Prot}).
\end{corol}
\begin{IEEEproof}
\iftoggle{Extended}{%
Please refer to Appendix \ref{App:Corol_Analytic}.
}{%
  The result in this corollary directly follows from the minimization of the dual function of a linear program.  Please refer to \cite[Appendix C]{GlobeCOM2015Arxiv} for the detailed
proof.
}
\end{IEEEproof}

Corollary~\ref{Corol:Analytic} provides an important intuition based on the statistics of the RF and FSO links. In particular, if the FSO link is statistically strong (cf. Case 1), the relay can forward all the information received from the users to the destination over the FSO backhaul link and the $\boldsymbol{\mathcal{R}}$-$\boldsymbol{\mathcal{D}}$ RF backhaul link remains inactive for all fading blocks, i.e., we obtain $\lambda^*=1$. However, if the statistical quality of the FSO link is not as strong e.g.  due to severe atmospheric conditions (cf. Case 2), the $\boldsymbol{\mathcal{R}}$-$\boldsymbol{\mathcal{D}}$ RF backhaul link becomes active in fading blocks which satisfy $\lambda C_1^{\mathrm{RF}}(\mathbf{H}_1) < [1-\lambda]C_2^{\mathrm{RF}}(\mathbf{H}_2) $. Thereby, the optimal value of $\lambda$ is obtained such that the sum of the average information rates sent from the relay to the destination over both the RF and the FSO links is identical to the average  information rate received at the relay from the users. 

\begin{remk}\label{Remk:Cap_General}
We note that the proposed problem formulation and the resulting protocol in Theorem~\ref{Theo:Opt_Prot} are provided in a general form such that coding schemes  different from those considered in this paper can be easily accommodated. In particular, using different coding schemes changes only the values of the instantaneous capacity expressions in (\ref{Eq:UR_Rate_RF}), (\ref{Eq:RU_Cap_RF}), and (\ref{Eq:RU_Cap_FSO}) such that the analysis of the proposed protocol and resource allocation policy remains valid. For instance, as explained in Section II.C., we assume that the users transmit with a priori fixed transmission rate in order to avoid the CSI feedback overhead from the relay to all the users. However, if this CSI overhead can be accommodated, then the users can transmit with adaptive rates in each fading block. Thereby, the maximum sum capacity of the multiple-access channel  is given by \cite{Cover}
\begin{IEEEeqnarray}{lll}\label{Eq:UR_Cap_RF}
  C_1^{\mathrm{RF}}[b] =   &\log_2\bigg\{ \Big| \mathbf{I}_J+\frac{1}{\sigma_{\boldsymbol{\mathcal{R}}}^2} \nonumber \\
  &\times \mathbf{H}_1[b] \mathrm{diag} \big\{P_{\boldsymbol{\mathcal{U}}}^{1},P_{\boldsymbol{\mathcal{U}}}^{2},\dots,P_{\boldsymbol{\mathcal{U}}}^{K} \big\} (\mathbf{H}_1[b])^H \Big| \bigg\}. \quad
\end{IEEEeqnarray}
\end{remk}

\section{Numerical Results}

The  values of the  parameters  for  the  RF  and  FSO  links used for the numerical results shown in this section are provided in Table~I. Moreover, we consider modified versions of the following two protocols as benchmark schemes \textit{i)} mixed RF/FSO relaying which does not employ the RF backhaul link \cite{GeorgeFSO} and \textit{ii)} conventional RF relaying \cite{Cover} which does not employ an FSO backhaul link. Unlike the original versions of these protocols, the relays of the benchmark schemes are also equipped with buffers. Hence, the nodes are able to transmit data for many consecutive fading blocks to average out the fading. Additionally, for conventional RF relaying, we optimize the ratio of the number of fading blocks allocated to the $\boldsymbol{\mathcal{U}}$-$\boldsymbol{\mathcal{R}}$ and $\boldsymbol{\mathcal{R}}$-$\boldsymbol{\mathcal{D}}$ RF links for throughput maximization, similar to \cite{Poor}. Furthermore, for all considered protocols, we assume that the relay employs the ZF detector to recover the information sent by the users.

\begin{table}
\label{Table:Parameter}
\caption{Values of the Numerical Parameters \cite{HybridRFFSOShi,HybridRFFSORobert}.\vspace{-0.2cm}} 
\begin{center}
\iftoggle{Extended}{%
\scalebox{0.45}
}{%
\scalebox{0.44}  
}
{
\begin{tabular}{|| c | c  | c ||}
  \hline
   \multicolumn{3}{||c||}{\textbf{RF Link}}\\ \hline \hline    
 Symbol & Definition & Value \\ \hline \hline
 $d^{k}$ & Distance between  user $k$ and the relay & $400$ m \\ \hline
   $P_{\boldsymbol{\mathcal{U}}}^{k}$ & Transmit power of user $k$ & $200$ mW ($23$ dBm) \\ \hline 
     $P_{\boldsymbol{\mathcal{R}}}^{\mathrm{RF}}$ & Relay transmit power over RF link & $2$ W ($33$ dBm) \\ \hline 
 $d$ & Distance between  the relay and the destination & $[1, 2]$ km \\ \hline
 $d^{\mathrm{RF}}_{\mathrm{ref}}$ & Reference distance of the RF link & $60$ m \\ \hline
 $(G^{\mathrm{RF}}_t,G^{\mathrm{RF}}_r)$ & Antenna gains for $\boldsymbol{\mathcal{U}}$-$\boldsymbol{\mathcal{R}}$ RF link & $(0,8)$ dBi \\ 
    & Antenna gains for $\boldsymbol{\mathcal{R}}$-$\boldsymbol{\mathcal{D}}$ RF link & $(10,15)$ dBi \\ \hline
  $N_0$ & Noise power spectral density at the RF receivers & $-114$ dBm/MHz \\ \hline 
  $N_F$ & Receiver noise figure of the RF receivers & $5$ dB  \\ \hline 
   $\lambda^{\mathrm{RF}}$ & Wavelength of RF signal & $85.7$ mm ($3.5$ GHz) \\ \hline 
      $W^{\mathrm{RF}}$ & Bandwidth of RF signal & $20$ MHz \\ \hline 
    $(\Omega,\Psi)$ & Rice  distribution parameters for $\boldsymbol{\mathcal{U}}$-$\boldsymbol{\mathcal{R}}$ RF link & $(0,1)$  \\ 
       & Rice  distribution parameters for $\boldsymbol{\mathcal{R}}$-$\boldsymbol{\mathcal{D}}$ RF link & $(4,1)$  \\ \hline  
    $\nu$ &  RF path-loss exponent & $3.5$ \\ \hline\hline 
       \multicolumn{3}{||c||}{\textbf{FSO Link}}\\ \hline \hline    
 Symbol & Definition & Value \\ \hline \hline
  $d$ & Distance between  the relay and the destination & $[1, 2]$ km \\ \hline
  $P_{\boldsymbol{\mathcal{R}}}^{\mathrm{FSO}}$ & Relay transmit power over FSO link & $40$ mW ($16$ dBm) \\ \hline 
  $\sigma^2$ & Noise variance at the FSO receivers & $10^{-14}$ $\mathrm{A}^2$ \\ \hline 
   $\lambda^{\mathrm{FSO}}$ & Wavelength of FSO signal & $1550$ nm ($193$ THz) \\ \hline
      $W^{\mathrm{FSO}}$ & Bandwidth of RF signal & $1$ GHz  \\ \hline   
       $\rho$ & Responsivity of FSO photo-detector & $0.5\frac{1}{\mathrm{V}}$   \\ \hline  
       $\kappa$ & Weather-dependent attenuation factor for &   $ 10^{-3}\times$\\ 
       & [clear air, haze, light fog, moderate fog, heavy fog]  & $[0.43,4.2,20,42.2,125]$  \\ \hline  
       $C_n^2$ & Weather-dependent index of refraction structure for &   \\ 
       & [clear air, haze, light fog, moderate fog, heavy fog]  & $ 10^{-15}\times[50,17,3,2,1]$  \\ \hline
       $\phi$ & Laser divergence angle  & $2$ mrad  \\ \hline  
       $r$ & Aperture radius  & $10$ cm  \\ \hline            
\end{tabular}
}
\end{center}
\vspace{-0.7cm}
\end{table}

\begin{figure}
\centering
\iftoggle{Extended}{%
\resizebox{1\linewidth}{!}{\psfragfig{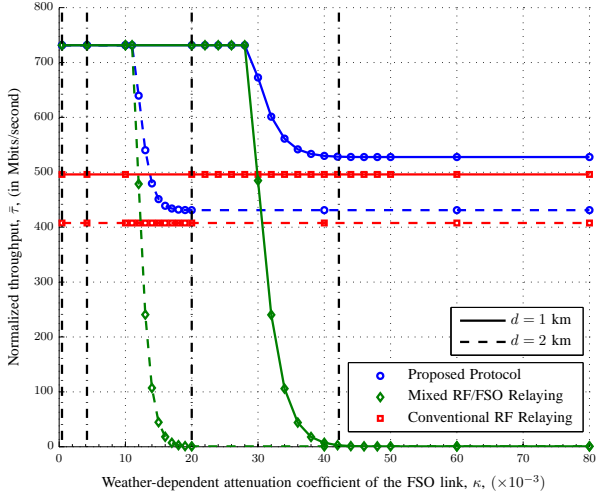}} 
}{%
\resizebox{1\linewidth}{!}{\psfragfig{Fig/Weather/Weather}}   
}\vspace{-1cm}
    \caption{Normalized throughput, $\bar{\tau}$, in Mbits/second vs. weather-dependent attenuation coefficient of the FSO link, $\kappa$, for $K=5$, $J=L=10$, $d=1$~km and $2$ km, and $R_{\boldsymbol{\mathcal{U}}}^{k}=8$ bits/symbol $\forall k$. The remaining parameters are given in Table~I. Moreover, from low to high values of $\kappa$, the vertical dashed-dotted lines represent the following weather conditions \cite{HybridRFFSORobert}: clear air, haze, light fog, and moderate fog, respectively.\vspace{-0.5cm}}
    \label{Fig:Weather}
\end{figure}

In Fig.~\ref{Fig:Weather}, we show the normalized achievable sum throughput of all users $\bar{\tau}=\frac{\tau}{T_{\boldsymbol{\mathcal{UR}}}}$ (in bits/second) versus the weather-dependent attenuation coefficient of the FSO link, $\kappa$, for $K=5$, $J=L=10$, $d=1$ km and $2$ km, and $R_{\boldsymbol{\mathcal{U}}}^{k}=8$ bits/symbol for $\forall k$. From Fig.~\ref{Fig:Weather}, we observe that for favorable atmospheric conditions, i.e., low values of $\kappa$, the relay is able to forward all the information received from the users to the destination over the FSO link. Hence, the $\boldsymbol{\mathcal{R}}$-$\boldsymbol{\mathcal{D}}$ RF backhaul link remains inactive in the proposed protocol, cf. Case~1 in Corollary~\ref{Corol:Analytic}. As a result, the achievable throughput of the proposed protocol and the mixed RF/FSO relaying protocol coincide. However, as $\kappa$ increases, i.e., the quality of the FSO link deteriorates due to severe atmospheric conditions, there is a critical value of $\kappa$ above which the FSO link is no longer able to forward all the user data, which is received at the relay, to the destination. Hence, the achievable throughput of the mixed RF/FSO relaying protocol decreases ultimately to zero for large values of $\kappa$. In contrast, although the achievable throughput of the proposed protocol also decreases due to the sharing of the RF bandwidth between the $\boldsymbol{\mathcal{U}}$-$\boldsymbol{\mathcal{R}}$ and $\boldsymbol{\mathcal{R}}$-$\boldsymbol{\mathcal{D}}$ RF links, it converges to a non-zero value for large values of $\kappa$, cf. Case~2 in Corollary~\ref{Corol:Analytic}. Note that the achievable throughput of the conventional RF relaying protocol does not depend on the quality of the FSO link. Moreover, for high values of $\kappa$ when the FSO link becomes unavailable, the proposed protocol still outperforms the conventional RF relaying protocol  due to the adaptive resource allocation based on the instantaneous CSI, cf. Theorem~\ref{Theo:Opt_Prot}. Finally, we observe that the critical value of $\kappa$ above which adaptive RF link selection becomes necessary depends on the system parameters. For instance, for the  parameters considered here and moderate foggy atmospheric conditions, i.e., $\kappa=20\times 10^{-3}$, the $\boldsymbol{\mathcal{R}}$-$\boldsymbol{\mathcal{D}}$ RF link is active if $d=2$ km but  inactive if  $d=1$ km.

\section{Conclusions}

A mixed RF and hybrid RF/FSO system was considered where multiple users transmit their data over an RF link to a relay node and the relay forwards the information to the destination over a hybrid RF/FSO backhaul link. The optimal resource allocation policy which shares the RF bandwidth resource between the RF links based on the CSI was derived. Our numerical results revealed the effectiveness of the proposed communication architecture and resource allocation policy even if the FSO link is affected by severe atmospheric conditions.

\iftoggle{Extended}{%

\appendices

\section{Proof of Theorem \ref{Theo:Opt_Prot}}\label{App:Theo_Opt_Prot}

In this appendix, our aim is to find the optimal resource allocation policy as a solution of the optimization problem given in (\ref{Eq:ProbUpp}). Since the cost function and the constraints in (\ref{Eq:ProbUpp}) are affline in the optimization variables $q[b],\,\,\forall b$, and $\tau^{\mathrm{upp}}$ and the feasible set is non-empty, Slater's condition is satisfied. Hence, the duality gap is zero \cite{Boyd}. Therefore, the  solution of the primal problem in (\ref{Eq:ProbUpp}) can be found from the solution of the dual problem of (\ref{Eq:ProbUpp}) \cite{Boyd}. Denoting the Lagrange multipliers corresponding to constraints  $\mathrm{C1}$ and $\mathrm{C2}$  by $\lambda_1$ and $\lambda_2$, respectively, the Lagrangian function corresponding to the optimization problem in (\ref{Eq:ProbUpp}) is obtained as
\begin{IEEEeqnarray}{ll}\label{Eq:LagFunction}
   \mathcal{L}(q[b],\tau^{\mathrm{upp}},\lambda_1,\lambda_2)  \,\,=  \tau^{\mathrm{upp}}  \\  + \lambda_1 \left[ \frac{1}{B} \sum_{b=1}^{B} q[b] N C_1^{\mathrm{RF}}[b]-\tau^{\mathrm{upp}}\right] \nonumber \\
   + \lambda_2\left[\frac{1}{B} \sum_{b=1}^{B} \big[ [1-q[b]] N C_2^{\mathrm{RF}}[b] + MN C^{\mathrm{FSO}}[b] \big ]-\tau^{\mathrm{upp}}\right]. \nonumber
\end{IEEEeqnarray}
The dual function is then given by
\begin{IEEEeqnarray}{ll}\label{Eq:DualFunction}
   \mathcal{D}(\lambda_1,\lambda_2)= \underset{q[b]\in[0,1],\tau^{\mathrm{upp}}\geq 0}{\mathrm{maximize}} \,\,\mathcal{L}(q[b],\tau^{\mathrm{upp}},\lambda_1,\lambda_2)
\end{IEEEeqnarray}
and the dual problem is given by
\begin{IEEEeqnarray}{ll}\label{Eq:DualProb}
    \underset{\lambda_1\geq 0,\lambda_2\geq 0}{\mathrm{minimize}} \,\,&\mathcal{D}(\lambda_1,\lambda_2).
\end{IEEEeqnarray}

To solve (\ref{Eq:ProbUpp}) using the dual problem in (\ref{Eq:DualProb}), we first obtain the primal variables $q[b]$ and $\tau^{\mathrm{upp}}$ for a given dual variables $\lambda_1$ and $\lambda_2$ from (\ref{Eq:DualFunction}). Then, the optimal $\lambda_1$ and $\lambda_2$ are obtained by solving the dual problem in (\ref{Eq:DualProb}).

\subsection{Optimal Primal Variables}

The optimal resource allocation variables, $q[b],\,\,\forall b$, and the optimal value of the upper bound, $\tau^{\mathrm{upp}}$,  are either  the boundary points of the feasible sets, i.e., $q[b]\in[0,1]$ and $\tau^{\mathrm{upp}}\geq 0$, or the stationary points which can be obtained by setting the derivatives
of the Lagrangian function in (\ref{Eq:LagFunction}) with respect to $\tau^{\mathrm{upp}}$ and $q[b]$ to zero. The derivatives
of the Lagrangian function in (\ref{Eq:LagFunction}) are given by
\begin{IEEEeqnarray}{rll}\label{Eq:StationaryMode}
    \frac{\partial\mathcal{L}}{\partial \tau^{\mathrm{upp}}} \,\,&= 1-\lambda_1-\lambda_2 \, \IEEEeqnarraynumspace \IEEEyesnumber \IEEEyessubnumber \\
    \frac{\partial\mathcal{L}}{\partial q[b]} \,\,&= \frac{1}{B} \left[\lambda_1 N C_1^{\mathrm{RF}}[b] - \lambda_2 N C_2^{\mathrm{RF}}[b] \right]. \quad  \IEEEyessubnumber 
\end{IEEEeqnarray}
If the derivative $\frac{\partial\mathcal{L}}{\partial \tau^{\mathrm{upp}}} $ is non-zero, the optimal value of $\tau^{\mathrm{upp}}$ is at the boundary of its feasible set, i.e., $\tau^{\mathrm{upp}}\to\infty$ or $\tau^{\mathrm{upp}}\to 0$, which cannot be the optimal solution. Therefore, the derivative $\frac{\partial\mathcal{L}}{\partial \tau^{\mathrm{upp}}} $ in  (\ref{Eq:StationaryMode}a) has to be zero which leads to $\lambda_1+\lambda_2 = 1$. Since the  probability density functions of the channel coefficients $\mathbf{H}_1$ and $\mathbf{H}_2$ are continuous, we obtain that  $\Pr\left\{\frac{\partial\mathcal{L}}{\partial q[b]}=0\right\}=0$ holds. Hence, the optimal value of $q[b]$ is always at the boundaries, i.e., 
\begin{IEEEeqnarray}{lll}\label{Eq:OptFixAppQFix}
   q^*[b] = {\begin{cases} 
    1,\quad &\mathrm{if}\quad \lambda_1 C_1^{\mathrm{RF}}[b] \geq \lambda_2 C_2^{\mathrm{RF}}[b]\\
    0,\quad &\mathrm{otherwise}
\end{cases}}
\end{IEEEeqnarray}
Defining $\lambda\triangleq \lambda_1=1-\lambda_2$, (\ref{Eq:OptFixAppQFix}) leads to the optimal link selection policy in (\ref{Eq:Opt_Prot}). 

\subsection{Optimal Dual Variable}

The optimal value of $\lambda$ is obtained by solving the dual problem in (\ref{Eq:DualProb}). In particular, substituting, the optimal values of $q^*[b]$ in (\ref{Eq:OptFixAppQFix}) for a given $\lambda$ into  (\ref{Eq:LagFunction}), we obtain the dual function in (\ref{Eq:DualFunction}). The dual function $\mathcal{D}(\lambda)$ is a linear function of $\lambda$ and its derivative is obtained as
\begin{IEEEeqnarray}{ll}\label{Eq:Dual_Deriv}
   \frac{\partial \mathcal{D}(\lambda)}{\partial \lambda}   =  &\frac{1}{B} \sum_{b=1}^{B} \Big[  q^*[b] N C_1^{\mathrm{RF}}[b] \nonumber \\ &- \big[(1-q^*[b]) N C_2^{\mathrm{RF}}[b] 
   + MN C^{\mathrm{FSO}}[b] \big ] \Big]. 
\end{IEEEeqnarray}
We note that for a given $\lambda$, the value of $\frac{\partial \mathcal{D}(\lambda)}{\partial \lambda}$ is fixed. Thereby, we can investigate the following mutually exclusive cases for the optimal $\lambda^*$:

\textit{Case 1:} If we assume that $\lambda^*>1$ holds, we obtain $q[b]=1,\,\,\forall b$ from (\ref{Eq:OptFixAppQFix}). Thereby, the derivative $\frac{\partial \mathcal{D}(\lambda)}{\partial \lambda}$ becomes positive and in order to minimize $ \mathcal{D}(\lambda)$, we have to decrease $\lambda$. In fact, for each $\lambda^*>1$ and $\epsilon>0$, we can find a new $\lambda=\lambda^*-\epsilon$ where $q[b]=1,\,\,\forall b$ still holds and the dual function is decreased. Hence, $\lambda^*>1$ cannot be optimal.

\textit{Case 2:} If we assume that $\lambda^*=0$ holds, we obtain $q[b]=0,\,\,\forall b$ from (\ref{Eq:OptFixAppQFix}). Intuitively, this cannot be the optimal dual variable since it leads to $\tau^{\mathrm{upp}}= 0$. In particular, the derivative $\frac{\partial \mathcal{D}(\lambda)}{\partial \lambda}$ becomes negative and in order to minimize $ \mathcal{D}(\lambda)$, we have to increase $\lambda$. In fact, we can assume a small $\lambda=\epsilon$ and obtain a positive $\tau^{\mathrm{upp}}$ and decrease the dual function. Hence, $\lambda^*=0$ cannot hold for the optimal solution.

\textit{Case 3:} Considering Case 1 and Case 2, $0<\lambda^*\leq 1$ has to hold. In order to find the optimal dual variable, we can employ the widely-adopted gradient  method \cite{BoydDual}.  The main idea is to minimize $\mathcal{D}(\lambda)$ by updating $\lambda$ along the gradient search direction. The updates are performed as 
\begin{IEEEeqnarray}{ll}
  \lambda[i+1] =  \left[\lambda[i]-\delta[i]\frac{\partial \mathcal{D}(\lambda)}{\partial \lambda}\bigg|_{\lambda=\lambda[i]}\right]_0^1.
  \end{IEEEeqnarray}
 This leads to the iterative algorithm in Theorem~\ref{Theo:Opt_Prot} with the updates given in (\ref{Eq:Update}).   The gradient method is guaranteed to converge to the optimal dual variable $\lambda^*$ provided that the step sizes $\delta[i]$ are chosen   sufficiently small \cite{BoydDual}. This completes the proof.

\section{Proof of Lemma~\ref{Lem:Achiev}}\label{App:Lem_Achiev} 
 
 The proof follows from a well-known property in queuing theory \cite{Neely}. Let us consider a buffer with stochastic arrival process $a[i]$, departure demand process (queue process) $d[i]$, and queue lengths $Q[i]$ in the $i$-th time instance. Thereby, the departure rate (queue service) is obtained as $b[i]=\min\{Q[i-1],d[i]\}$ and  $Q[i]$ evolves as $Q[i]=\max\{Q[i-1]-d[i],0\}+a[i]$. However, the average departure rate $\mathbbmss{E}\{b\}$ (averaged over the time slots) can be expressed independently from the dynamics of the queue as 
\begin{IEEEeqnarray}{lll}\label{MinRateBound}
   \mathbbmss{E}\{b\} \,\,&= \begin{cases}
   \mathbbmss{E}\{d\}, \,\, &\mathrm{if}\,\, \mathbbmss{E}\{a\}>\mathbbmss{E}\{d\} \\
\mathbbmss{E}\{a\}=\mathbbmss{E}\{d\}, &\mathrm{if}\,\,\mathbbmss{E}\{a\}=\mathbbmss{E}\{d\} \\
\mathbbmss{E}\{a\}, \,\, &\mathrm{if}\,\, \mathbbmss{E}\{a\}<\mathbbmss{E}\{d\}
\end{cases}
\end{IEEEeqnarray}
In particular, if  $\mathbbmss{E}\{a\}>\mathbbmss{E}\{d\}$ holds, i.e., the average rate flowing into the buffer is larger than  the average capacity of the respective departure channel, then, as the number of time instances grow to infinity, the buffer  always has enough information to supply because the amount of information in the queue increases over time and we obtain  $\mathbbmss{E}\{b\}=\mathbbmss{E}\{d\}$. On the other hand, if $\mathbbmss{E}\{a\}<\mathbbmss{E}\{d\}$ holds, i.e., the average information flowing into the buffer is less than the average capacity of the respective departure channel, then by the law of conservation of flow, we obtain $\mathbbmss{E}\{b\}=\mathbbmss{E}\{a\}$. If  $\mathbbmss{E}\{a\}=\mathbbmss{E}\{d\}$ holds, we obtain $\mathbbmss{E}\{b\}=\mathbbmss{E}\{a\}=\mathbbmss{E}\{d\}$ due to the continuity property. Hence, from (\ref{MinRateBound}), we can conclude that $\mathbbmss{E}\{b\}=\min\{\mathbbmss{E}\{a\},\mathbbmss{E}\{d\}\}$ holds which leads to the expression given in (\ref{Eq:OptUpp}) for the upper bound, i.e., $\tau=\tau^{\mathrm{upp}}$. This completes the proof.

 \section{Proof of Corollary~\ref{Corol:Analytic}}\label{App:Corol_Analytic}
 
 If $\mathbbmss{E}\big\{C_1^{\mathrm{RF}}(\mathbf{H}_1) \big\} \leq M \mathbbmss{E}\big\{C^{\mathrm{FSO}}(g) \big\}$ holds, from (\ref{Eq:Dual_Deriv}), we obtain that $\frac{\partial \mathcal{D}(\lambda)}{\partial \lambda}<0$ holds regardless of the value of $\lambda$. Hence, in order to minimize the dual function $\mathcal{D}(\lambda)$, we have to increase $\lambda$. Moreover, since $\lambda\in(0,1]$ holds, we obtain $\lambda^*=1$. On the other hand, if $\mathbbmss{E}\big\{C_1^{\mathrm{RF}}(\mathbf{H}_1) \big\} > M \mathbbmss{E}\big\{C_1^{\mathrm{FSO}}(g) \big\}$ holds, we can conclude that $0<\lambda^*<1$ and its value is unique. In other words, according to (\ref{Eq:OptUpp}), $ \tau^{\mathrm{upp}} =N  \min\big\{\bar{C}_1^{\mathrm{RF}},\bar{C}_2^{\mathrm{RF}}+\bar{C}^{\mathrm{FSO}}\big\}$ where $\bar{C}_1^{\mathrm{RF}}$ is a monotonically increasing function of $\lambda$, $\bar{C}_2^{\mathrm{RF}}$ is a monotonically decreasing function of $\lambda$, and $\bar{C}^{\mathrm{FSO}}$ does not depend on $\lambda$. Therefore, the optimal $\lambda$ has to be chosen such that $\bar{C}_1^{\mathrm{RF}}=\bar{C}_2^{\mathrm{RF}}+\bar{C}^{\mathrm{FSO}}$ holds which leads to (\ref{Eq:Eq_Opt_Lam}). This completes the proof.

}{%
  
}

\section*{Acknowledgment}
This publication was made possible by the NPRP award [NPRP 5-157-2-051] from the Qatar National Research Fund (a member of the Qatar Foundation). The statements made herein are solely the responsibility of the authors.

\bibliographystyle{IEEEtran}
\bibliography{Ref_19_02_2015}
\end{document}